\documentclass[aps,preprint,superscriptaddress]{revtex4-1}
\usepackage{graphicx}  
\usepackage{dcolumn}   
\usepackage{bm}        
\usepackage{amssymb}   
\usepackage{amsmath}
\usepackage{xcolor}
\usepackage{appendix}

\begin{document}

\title{Spatiotemporal control of laser intensity through cross-phase modulation}

\author{T.T. Simpson}
\email{tsim@lle.rochester.edu} 
\affiliation{
Laboratory for Laser Energetics, University of Rochester, Rochester, NY 14623, USA}
\author{D. Ramsey}
\affiliation{
Laboratory for Laser Energetics, University of Rochester, Rochester, NY 14623, USA}
\author{P. Franke}
\affiliation{
Laboratory for Laser Energetics, University of Rochester, Rochester, NY 14623, USA}
\author{K. Weichman}
\affiliation{
Laboratory for Laser Energetics, University of Rochester, Rochester, NY 14623, USA}
\author{M.V. Ambat}
\affiliation{
Laboratory for Laser Energetics, University of Rochester, Rochester, NY 14623, USA}
\author{D. Turnbull}
\affiliation{
Laboratory for Laser Energetics, University of Rochester, Rochester, NY 14623, USA}
\author{D.H. Froula}
\affiliation{
Laboratory for Laser Energetics, University of Rochester, Rochester, NY 14623, USA}
\author{J.P. Palastro}
\email{jpal@lle.rochester.edu} 
\affiliation{
Laboratory for Laser Energetics, University of Rochester, Rochester, NY 14623, USA}
\date{\today}

\email{tsim@lle.rochester.edu} 
\email{jpal@lle.rochester.edu} 



\begin{abstract}
Spatiotemporal pulse shaping provides control over the trajectory and range of an intensity peak. While this control can enhance laser-based applications, the optical configurations required for shaping the pulse can constrain the transverse or temporal profile, duration, or orbital angular momentum (OAM). Here we present a novel technique for spatiotemporal control that mitigates these constraints by using a “stencil” pulse to spatiotemporally structure a second, primary pulse through cross-phase modulation (XPM) in a Kerr lens. The temporally shaped stencil pulse induces a time-dependent focusing phase within the primary pulse. This technique, the “flying focus X,” allows the primary pulse to have any profile or OAM, expanding the flexibility of spatiotemporal pulse shaping for laser-based applications. As an example, simulations show that the flying focus X can deliver an arbitrary-velocity, variable-duration intensity peak with OAM over distances much longer than a Rayleigh range.
\end{abstract}

\maketitle

\section{Introduction}

Spatiotemporal pulse shaping offers a new paradigm for controlling laser intensity. By exploiting space-time correlations in the amplitude or phase of a laser pulse, a number of recent techniques have created arbitrary-velocity intensity peaks that remain nearly propagation invariant over distances much longer than a Rayleigh range \cite{kondakci2017diffraction,sainte2017controlling,froula2018spatiotemporal,Kondakci2019,li2020velocity,palastro2020dephasingless,caizergues2020phase,JollyOpticsExpress}. These features promise to revolutionize a wide range of laser-based applications, including high-power amplifiers \cite{turnbull2018raman}, radiation sources \cite{howard2019photon, franke2021optical, ramsey2021nonlinear}, and compact accelerators \cite{DebusLWFA,palastro2020dephasingless,caizergues2020phase,ramsey2020vacuum,palastro2021laser}. Nevertheless, each of the existing techniques requires an optical configuration that constrains properties of the intensity peak, such as its transverse or temporal profile, duration, or orbital angular momentum (OAM). As an example, the original, ``chromatic" flying focus uses a chirp and a chromatic lens to control the time and location at which each temporal slice of a laser pulse comes to focus, respectively \cite{froula2018spatiotemporal, sainte2017controlling}. While this technique offers some flexibility to shape the transverse profile of the far-field, the chromatic aberration and chirp can place a lower bound on the duration of the intensity peak that is much larger than the transform-limited duration \cite{sainte2017controlling,palastro2018ionization,froulaPOP}. An alternative technique, the ``achromatic" flying focus, employs the spherical aberration of an axiparabola \cite{smartsev2019axiparabola,oubrerie2021axiparabola} to focus different annuli in the near-field to different axial locations in the far-field and an echelon to adjust their relative timing \cite{palastro2020dephasingless}. Here, the intensity peak can have a near transform-limited duration, but the flattop transverse profile required in the near-field and the spherical aberration of the axiparabola fully determine the far-field profile. 

Both the chromatic and achromatic flying focus techniques use linear optical elements to structure a pulse with space-time correlations, but nonlinear processes, such as self-phase modulation, can also produce space-time correlations \cite{marburger1975self,shen1975self,brodeur1997moving,couairon2007femtosecond,simpson2020nonlinear}. Self-phase modulation refers to the phase accrued by a laser pulse as it traverses a nonlinear medium and modifies the refractive index. In a Kerr nonlinear medium, the change in the refractive index ($\delta n$) is proportional to the intensity ($I$): $\delta n = n_2I$, where $n_2$ is the nonlinear refractive index and linear polarization is assumed throughout. When $n_2>0$, the phase velocity is reduced in regions of high intensity, creating a wavefront curvature that can counteract diffraction, i.e., the pulse self-focuses \cite{kelley1965self,marburger1975self,shen1975self, sun1987self}. At sufficiently high power, this curvature can even overcome diffraction, resulting in the transverse collapse of the pulse until the intensity reaches a threshold for activating an arrest mechanism \cite{fibich1996small,bang2002collapse,kosareva2011arrest,couairon2007femtosecond}. The recently described ``self-flying focus" combines this effect with temporal pulse shaping to control the trajectory of an intensity peak in a Kerr medium \cite{simpson2020nonlinear}. This approach can deliver an arbitrary velocity intensity peak over distances comparable to the focal length, but nonlinear evolution of the transverse and temporal profiles during self-focusing collapse and arrest significantly complicate any attempt at far-field shaping. Further, the arrest mechanism, e.g., ionization, can irreversibly damage the medium. 

Cross-phase modulation (XPM) refers to the phase that one laser pulse induces onto another as they copropagate through a nonlinear medium \cite{agrawal2000nonlinear,boyd2019nonlinear}. For two pulses ($P$ and $S$ for specificity) overlapped in a Kerr medium, the change in refractive index experienced by pulse $P$ ($\delta n_P$) will depend on both its intensity ($I_P$) and that of pulse $S$ ($I_S$): $\delta n_P = n_2(I_P + \alpha I_S)$, where $\alpha$ depends on their relative polarizations. As a consequence, the phase velocity of pulse $P$ can be lower in regions where either pulse has a high intensity. If $I_S\gg I_P$, the intensity profile of pulse $S$ will dictate the local phase velocity and resulting wavefront curvature acquired by pulse $P$. That is, a high-intensity pulse can induce the focusing of a colocated low-intensity pulse \cite{mckinstrie1988nonlinear, agrawal1990induced,HafiziWater}. 

Here we describe a novel technique for spatiotemporal control: the ``flying focus X," which combines temporal pulse shaping with XPM to produce an ultrashort, arbitrary velocity intensity peak, with or without OAM, over distances far greater than a Rayleigh range. Specifically, a temporally shaped, high-intensity ``stencil” pulse induces the time-dependent focusing of a second, primary pulse through XPM in a Kerr lens [Fig. 1].  The minimum and maximum intensity of the stencil pulse ($S$) set the focal range of the primary pulse ($P$), while its duration sets the velocity of the resulting intensity peak. Use of a stencil pulse mitigates the constraints on the primary pulse, allowing for spatiotemporal control independent of properties such as the far-field duration, transverse profile, or OAM. In effect, these constraints are offloaded onto the stencil pulse. As a result, the flying focus X provides unprecedented flexibility for structuring the far-field properties of the intensity peak, promising to further enable or enhance a wide range of laser-based applications. 

\begin{figure}
\centering\includegraphics[width=\textwidth]{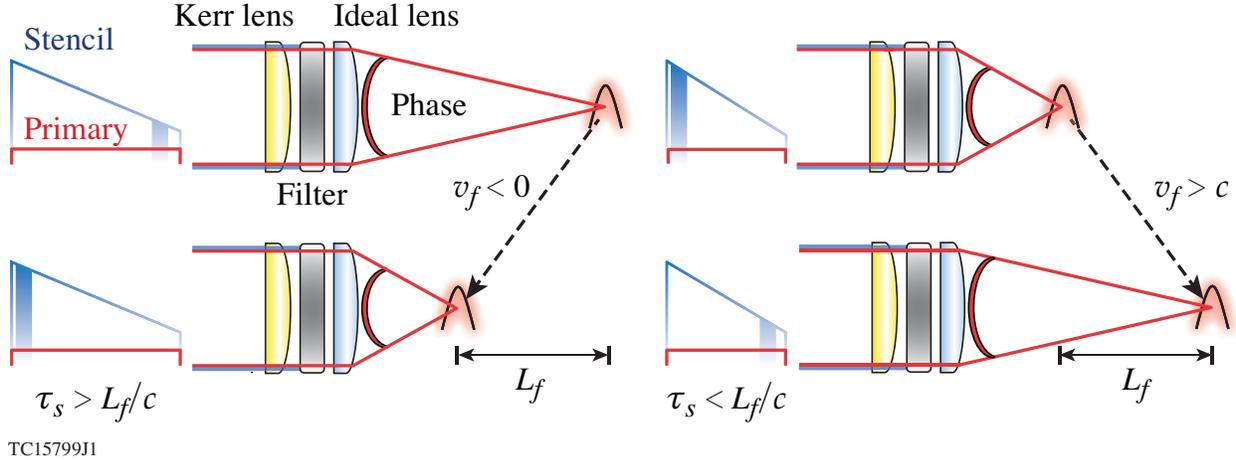}
\caption{Schematic of the flying focus X. A high-intensity stencil pulse with a flattop transverse and shaped temporal profile co-propagates with a low-intensity primary pulse. The stencil induces a time-dependent focusing phase on the primary in a parabolic Kerr lens. A filter removes the stencil, while an ideal lens provides additional focusing for the primary. The resulting focus travels over a distance $L_f$ at a velocity $v_f$, both of which can be tuned by shaping the temporal profile of the stencil. A stencil pulse with an intensity that ramps up in time can produce either negative (left) or positive superluminal (right) focal velocities, depending on its duration ($\tau_S$).}
\end{figure}

\section{The Flying Focus X}
Figure 1 illustrates how temporal pulse shaping and XPM can be used to control the trajectory and range of an intensity peak. A stencil pulse and a primary pulse, overlapped in space and time, co-propagate along the z axis and enter a convex, parabolic Kerr lens. The two pulses have distinct wavelengths or polarization vectors and $I_S \gg I_P$. The stencil pulse has a flattop transverse profile, which, in combination with the parabolic shape of the Kerr lens, ensures that the primary pulse acquires an aberration-free wavefront curvature. The instantaneous intensity of the stencil dictates the time dependence of that curvature. Specifically, the primary pulse exits the Kerr lens with the phase
\begin{equation}
    \phi_K(r,\tau)=-\frac{\alpha k_0 r^2 n_2 I_S(\tau)}{2R},
\end{equation}
where $\tau = t - z/c$ denotes time within the primary pulse, $k_0$ its vacuum wavenumber, and $R$ the Kerr lens radius of curvature (see Appendix I). The phase ($\phi_K$) is equivalent to that applied by an ideal lens, but with a time-dependent focal length: $f_K(\tau)=R/\alpha n_2I_s(\tau)$. After the Kerr lens, a wavelength or polarization filter removes the stencil pulse. The primary pulse continues through a second, linear lens such that its total phase ($\phi_P$) becomes $\phi_P = \phi_K + \phi_l$, where $\phi_l = -k_0r^2/2f_l$ and $f_l$ is the linear lens focal length. 

Upon exiting the linear lens, each time slice within the primary pulse will have a focal length determined by its total wavefront curvature:
\begin{equation}
    f(\tau)=f_l\Bigl[1+\frac{\alpha f_ln_2I_S(\tau)}{R}\Bigr]^{-1}.
\end{equation}
A time slice that was overlapped with a low stencil intensity in the Kerr lens will focus farther from the optical assembly and closer to the linear lens focal point; a time slice that was overlapped with a high stencil intensity will focus closer to the optical assembly and farther from the linear lens focal point. Over the duration of the primary pulse, the focal point will move within a range ($L_f$) defined by the minimum and maximum intensity of the stencil pulse:
\begin{equation}
    \frac{L_f}{f_l}=\frac{1}{1+\frac{\alpha f_ln_2}{R}I_{S,\mathrm{min}}} - \frac{1}{1+\frac{\alpha f_ln_2}{R}I_{S,\mathrm{max}}}. 
\end{equation}
For typical parameters, the linear lens has a smaller focal length (greater focusing power) than the Kerr lens, i.e., $f_l \ll f_K$. Nevertheless, the Kerr lens can still create a focal range that is a significant fraction of the linear lens focal length: $L_f \approx (\alpha f_l^2n_2/R)(I_{S,\mathrm{max}}-I_{S,\mathrm{min}})$.

By appropriately shaping the temporal profile of the stencil pulse, the focal point can be made to move through the focal region with a desired trajectory. Each time slice within the primary pulse will focus at a time $t_f = \tau + f(\tau)/c$. A general focal trajectory, $f(t_f)$, can be found by iterating to solve for $f(\tau)$. For a focal point moving at a constant velocity ($v_f$), $f(\tau) = f_0 + u\tau$, where $f_0$ is the focal point of the initial time slice and $u = cv_f/(c-v_f)$ is the reduced velocity. Using this expression for $f(\tau)$, one can invert Eq. (2) to find the required temporal profile of the stencil pulse:
\begin{equation}
I_S(\tau)=\frac{R}{\alpha f_ln_2}\Bigl(\frac{f_l}{f_0 + u\tau} -1\Bigr).
\end{equation}
For either a superluminal ($v_f>c$) or backwards ($v_f < 0$) focal velocity, $u<0$ and the stencil intensity must ramp up in time [Fig. 2(a)]. For a subluminal focal velocity ($0<v_f<c$), $u>0$ and the stencil intensity must ramp down in time. The total duration of the stencil and primary pulses ($\tau_S$) depends only on the focal range and the reduced velocity, i.e., $\tau_S = L_f/|u|$. Because the focal range ($L_f$) is typically much smaller than the linear lens focal length ($f_l$), the stencil profile can often be approximated as a simple linear ramp: $I_S(\tau) \approx (R/\alpha f_0n_2)(1 - f_0/f_l - u\tau/f_0)$.

From the stencil pulse duration, the focal velocity can be expressed in terms of physical parameters of the flying focus X: $v_f/c = (1\mp c\tau_S/L_f)^{-1}$, where the minus and plus signs correspond to ramp-up ($u<0$) and ramp-down ($u>0$) pulses, respectively [Fig. 2(b)]. This illustrates two points. First, for a desired focal range, the focal velocity can be tuned by adjusting the stencil pulse duration. Second, the transition from a superluminal to backwards travelling focus occurs when the pulse duration of a ramp-up pulse ($u<0$) becomes longer than the focal range (i.e., $\tau_S>L_f/c$). 
\begin{figure}
\centering\includegraphics[width=\textwidth]{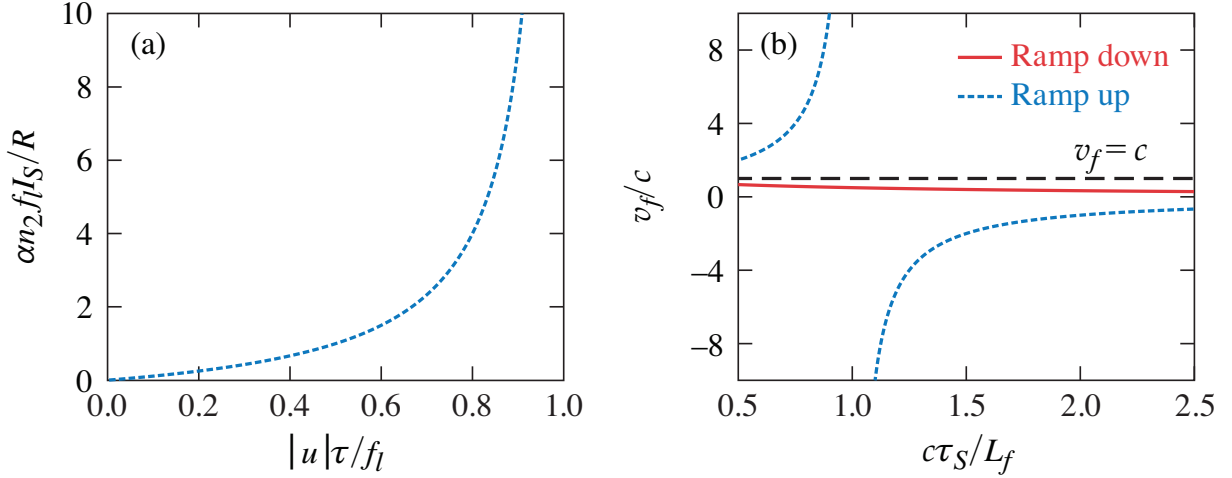}
\caption{Design space for a flying focus X pulse. (a) The intensity profile for a ramp-up ($u<0$) stencil pulse as a function of time. Starting at $\tau=0$ represents a pulse with $f_0 = f_l$; starting at a time $\tau=\Delta \tau$ represents a pulse with $f_0 = f_l - L_f\Delta \tau/\tau_S$. The intensity profile for a ramp-down pulse ($u>0$) corresponds to flipping the horizontal axis and appropriately shifting in time for a particular $f_0$. (b) The focal velocity as a function of the stencil pulse duration and focal range.}
\end{figure}

The primary pulse can have any transverse profile. Because the combined Kerr and linear lenses create an ideal parabolic wavefront, a natural choice is a Laguerre-Gaussian (LG) mode, which, in addition to allowing for OAM, has well-known expressions for its spot size, radius of curvature, and Gouy phase. With this choice, the spot size for each time slice within the primary pulse is given by $w_P(\tau)=w_0[(1-z/f(\tau))^2 + (z/Z_0)^2 ]^{1/2}$, where $w_0$ is the spot size in the near-field ($z=0$) and $Z_0 = k_0w_0^2/2$. From the spot size and conservation of power in each time slice of the primary pulse, one can derive the spatiotemporal profile of the intensity peak
\begin{equation}
I_f(z,t)=\Bigl[ 1 + \frac{(t-z/v_f + f_0/u)^2}{\tau_f^2} \Bigr] ^{-1}\frac{w_0^2}{w_f^2}I_{P0}(\tau),
\end{equation}
where $I_{P0}(\tau)$ is the intensity profile of the primary pulse in the near-field,
\begin{equation}
\tau_f = \frac{|c-v_f|}{|cv_f|}Z_{R}
\end{equation}
is the duration of the intensity peak, $w_f = 2f_l/k_0w_0$ is its spot size, $Z_R = k_0w_f^2/2$ is the Rayleigh range, and $L_f \ll f_l$ has been assumed. Note that each LG mode can have its peak intensity at a different radius; Eq. (5) is the temporal intensity profile at that radius.

The duration of the intensity peak ($\tau_f$) can be interpreted as the rate at which adjacent time slices within the primary pulse come in and out of focus, i.e., $Z_R/|u|$. The duration can be tuned by adjusting the focal geometry: stronger (weaker) focusing shortens (lengthens) the Rayleigh range, which decreases (increases) the duration. In general, the intensity peak can have an ultrashort duration that is much smaller than that of the primary pulse. The time-dependent refractive index experienced by the primary pulse in the Kerr lens provides the bandwidth necessary to support the ultrashort duration of the intensity peak in the far-field. The total bandwidth acquired by the primary pulse can be calculated using the linear ramp approximation for the stencil profile and the Kerr lens phase (Eq. (1)): $\Delta \omega = 2\partial_{\tau} \phi_K |_{r=w_0}$, or $\Delta \omega = 2 |u|/Z_R \propto \tau_f^{-1}$,  consistent with Eq. (6). Note that the required bandwidth increases rapidly as $v_f$ approaches $c$. As an example, with $v_f = 1.01\,c$ and $Z_R = 300\,\mathrm{\mu m}$, $\tau_f = 10 \,\mathrm{fs}$.

In order for the intensity peak to have a constant maximum throughout the focal region, the primary pulse must have a flattop intensity profile in time with a duration equal to that of the stencil pulse ($\tau_S$). By shaping the temporal intensity profile of the primary pulse ($I_{P0}$), the intensity peak can also exhibit more exotic patterns. For instance, modulating the temporal profile of the primary pulse will create an oscillating intensity distribution along the optical axis, similar to Ref. \cite{palastro2018ionization}. 

Designing a flying focus X pulse for an experiment or application follows a straightforward procedure. Several laser-based applications require (1) sustaining a high intensity over an extended distance and (2) that the velocity of the intensity peak conform to some process \cite{hebling2002velocity,DamicoTHz,kim2008coherent,rundquist1998phase,DurfeeHHG,popmintchev2012bright,durfee1993light,milchberg1996development,tajima1979laser,esarey2009physics,malkin1999fast,trines2011simulations,wilks1988frequency,dias1997experimental}. The first step is then to specify the focal range ($L_f$) and the focal velocity ($v_f$). Next, one would set the focal geometry ($f_l$) based on either a desired far-field spot size ($w_f$) or intensity peak duration ($\tau_f$). As demonstrated by Eq. (6), these two properties are interdependent. However, for a fixed $\tau_f$, a larger root-mean-square spot size can be achieved by using higher-order Laguerre-Gaussian modes. Finally, one would select a Kerr lens material with a large $n_2$, low absorption and weak nonlinear dispersion at the primary and stencil wavelengths, and the smallest possible radius of curvature. At optical frequencies, for instance, ZnS is an excellent candidate \cite{sheik1991dispersion}. With all other parameters specified, the duration and profile of the stencil pulse can be calculated from $\tau_S = (L_f/c)|c/v_f-1|$ [Fig 2(b)] and Eq. (2), respectively. Note that while the previous discussion considered a positive Kerr lens, the flying focus X can also be designed with a negative Kerr lens (see Appendix II). 

\begin{table}
\caption{Near-field parameters for the two simulated flying focus X designs. Note that only the properties of the stencil pulse change, illustrating the versatility of a single optical configuration to accommodate different focal velocities and transverse profiles. For both examples, $\lambda_0 = 2\pi/k_0 = 1.054 \,\mu \text{m}$, $L_f = 1\,\text{cm} = 33 \,Z_R$, and $w_f = 10 \,\mu \text{m}$. ZnS was used for the Kerr lens and the stencil and primary pulses had parallel polarizations, such that $\alpha = 2$ \cite{sheik1991dispersion,agrawal2000nonlinear}. }
\begin{ruledtabular}
\begin{tabular}{lcc}
$ $ & $v_f = 1.01\,c$ & $v_f = -c$ \\
\hline
$\ell$ & 0 & 1\\
$\tau_S$ (ps) & $0.33$ & $67$\\
$f_l$ (cm) & 60 & 60 \\
$w_0$ (cm) & 2.0  & 2.0 \\
$n_2$ (cm$^2$/W) & $8.4\times10^{-15}$ & $8.4\times10^{-15}$\\
$R$ (cm) & 1 & 1\\
$I_{S,\mathrm{max}}$ (W/cm$^2$) & $1.7\times10^{10}$ & $1.7\times10^{10}$
\end{tabular}
\end{ruledtabular}
\label{tab:t1}
\end{table}

\section{Simulation Results}
To demonstrate the versatility of the flying focus X, simulations (see Appendix III for details) were conducted for two focal velocities, each with relevance to laser-based applications (Table I). The transverse structure of the primary pulse in each case was an LG$_{\ell,p}$ mode, where $\ell$ and $p$ denote the OAM and radial quantum numbers, respectively. The first example features a superluminal ($v_f = 1.01 c$) intensity peak with $p=\ell=0$, i.e., a standard transverse Gaussian profile, ideal for compact accelerators (Fig. 3) \cite{esarey2009physics,palastro2020dephasingless,palastro2021laser}. The second example features a backwards ($v_f=-c$) travelling intensity peak with $p=0$ and $\ell = 1$, i.e., a ``donut" mode, ideal for exotic high-power amplification and photon acceleration schemes (Fig. 4) \cite{JorgeOAMRaman,franke2021optical}.
\begin{figure}
\centering\includegraphics[width=0.8\textwidth]{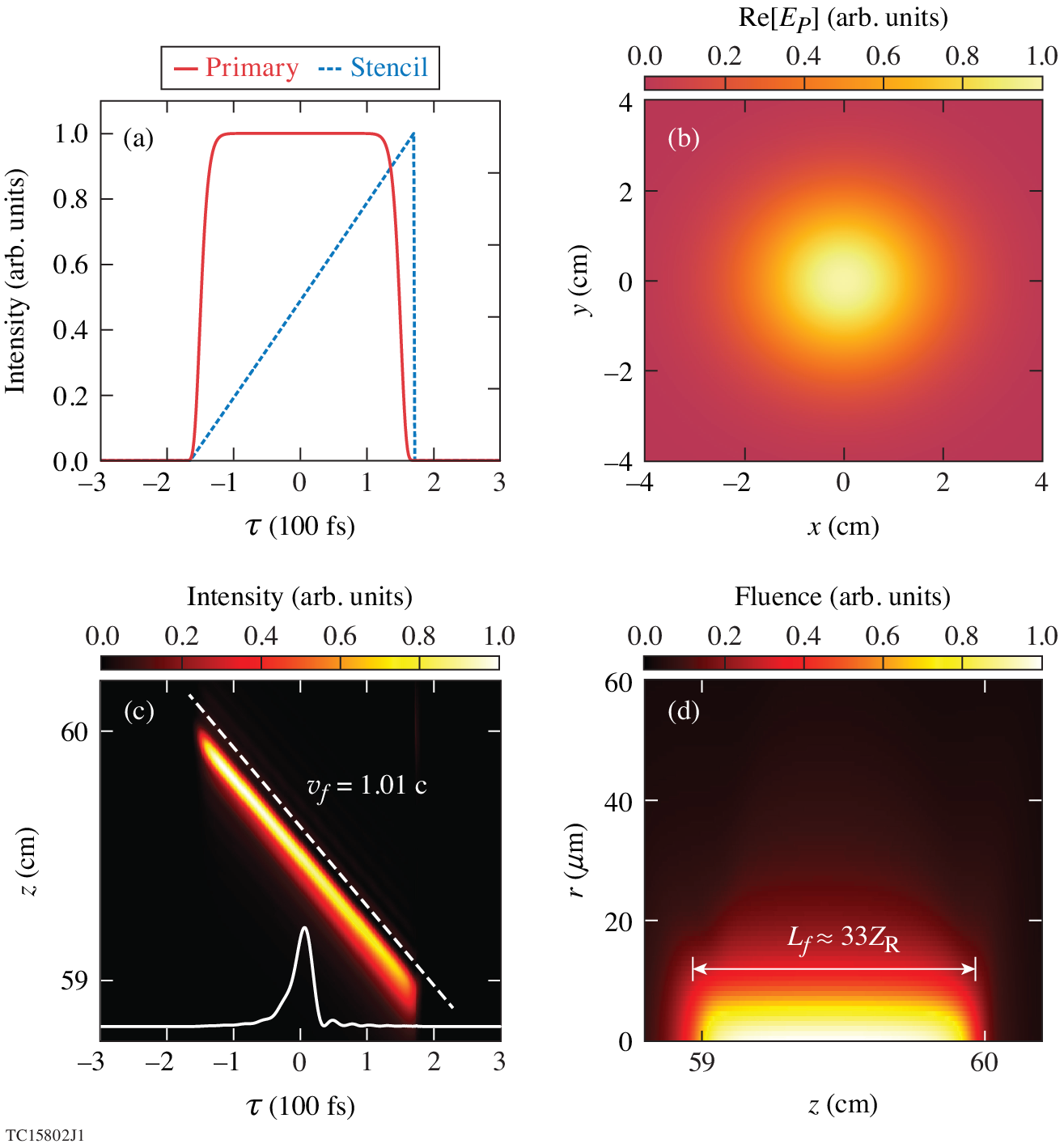}
\caption{Simulation results for an LG$_{0,0}$ flying focus with $v_f = 1.01 c$, demonstrating velocity control, an ultrashort duration, and near propagation invariance through the focal region. (a) The temporal profile of the stencil and primary pulses at the entrance of the Kerr lens. (b) The transverse electric field of the primary pulse at the entrance of the Kerr lens. (c) The on-axis intensity of the primary pulse in the far-field and a lineout of its temporal profile. (d) The fluence of the primary pulse in the far-field. All values are normalized to their respective maxima.}
\end{figure}
\begin{figure}
\centering\includegraphics[width=0.8\textwidth]{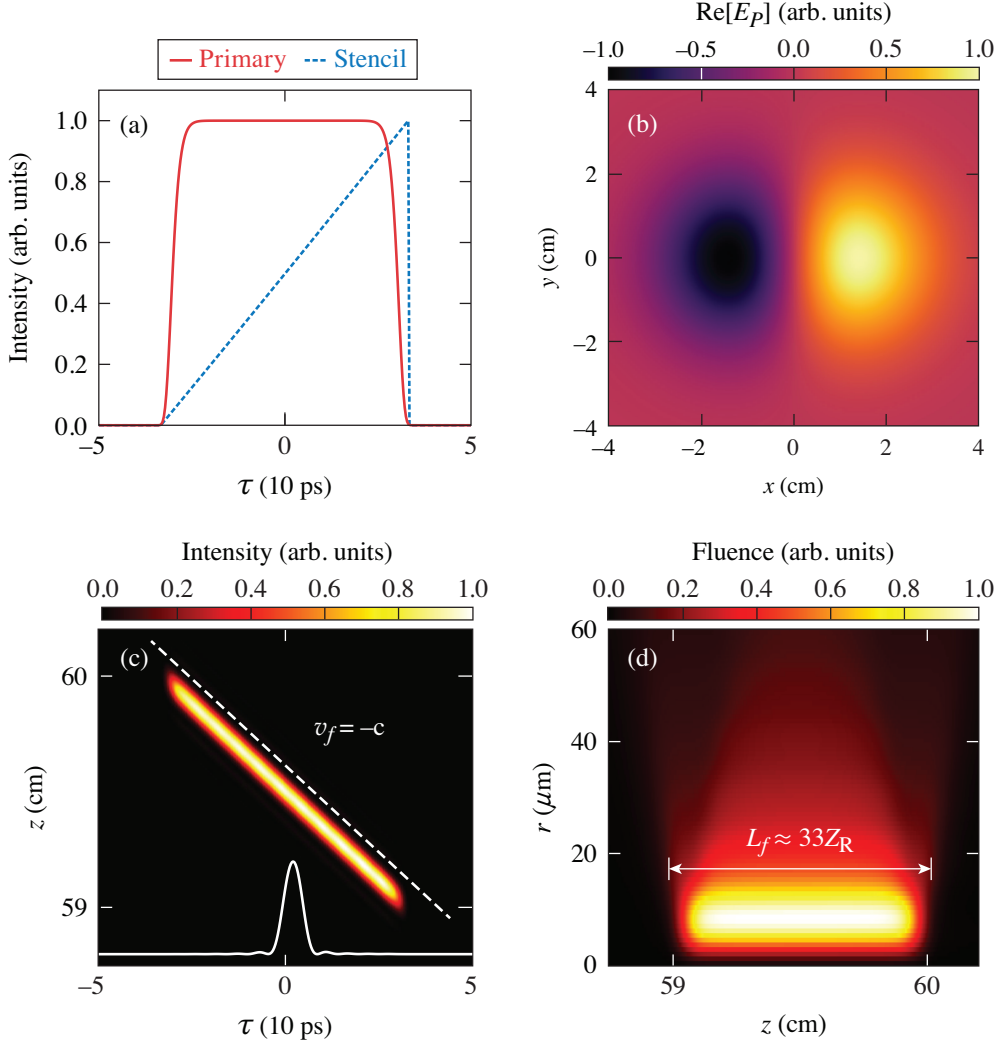}
\caption{Simulation results for an LG$_{0,1}$ flying focus with $v_f = -c$, demonstrating velocity control, a short duration, and near propagation invariance through the focal region. (a) The temporal profile of the stencil and primary pulses at the entrance of the Kerr lens. (b) The transverse electric field of the primary pulse at the entrance of the Kerr lens. (c) The intensity of the primary pulse in the far-field at the radial location of its maximum intensity ($r = w_f/\sqrt{2}$)  and a lineout of its temporal profile. (d) The fluence of the primary pulse in the far-field. All values are normalized to their respective maxima.}
\end{figure}

Figure 3 illustrates that the flying focus X can create an ultrashort-duration intensity peak traveling near the vacuum speed of light over many Rayleigh ranges.  As described by Eq. (4), the intensity of the stencil pulse ramps up linearly in time. The primary pulse has a super-Gaussian (SG20) temporal profile to ensure a constant intensity in the far-field. Throughout the focal region, spanning $z=59\,\text{cm}$ to $z=60\,\text{cm}$, the intensity peak remains ultrashort, with a duration comparable to the value predicted by Eq. (6), i.e., $\tau_f = 10\,\text{fs}$. The intensity peak maintains its transverse Gaussian profile with a diffraction-limited spot over the entire focal range. 

Figure 4 illustrates that the flying focus X can create an intensity peak with OAM that counterpropagates with respect to its phase fronts over many Rayleigh ranges. The stencil and primary pulses have the same temporal shape as in Fig. 3, but have much longer durations to accommodate the negative focal velocity [Fig. 2(b)]. As predicted by Eq. (6), the negative focal velocity results in the intensity peak having a longer effective duration, $\tau_{f}\approx 2\,\text{ps}$. The intensity peak maintains its OAM profile with a diffraction-limited spot over the entire focal range. 

\section{Discussion}

While the flying focus X offers additional flexibility to shape the far-field profile of the intensity peak, it does present some trade-offs. Foremost, the stencil pulse is filtered out and discarded after the Kerr lens, making the design inherently inefficient. This is exacerbated by the fact that the stencil pulse must be much more intense than the primary pulse to ensure that the induced focusing is not disrupted by self-phase modulation (i.e., $I_S \gg I_P/2$). The inefficiency can be mitigated by setting $I_{S,\text{min}}=0$, which minimizes the total energy of the stencil pulse, but fixes the maximum focal location to $f=f_l$. The constraint on the relative intensities can be eliminated if both the stencil and the primary pulses have a flattop, transverse intensity profile. From the resulting Kerr lens phase, i.e., $\phi_K(r,\tau)=-k_0n_2[\alpha I_S(\tau) + I_P]r^2/2R$, the temporal profile of the stencil can be shaped to accommodate SPM for any primary intensity, while still providing control over the far-field trajectory. Specifically, $I_S(\tau)=\frac{R}{\alpha f_ln_2}(\frac{f_l}{f_0 + u\tau} -1)-\frac{1}{\alpha}I_P$. Note, however, that this approach limits the far-field transverse profile to an Airy pattern and precludes a primary pulse with OAM. To eliminate the constraint on the relative intensities for a primary pulse with OAM, the shape of the Kerr lens can be changed to compensate the transverse spatial profile of the OAM mode. 

In addition to its inefficiency, the flying focus X requires stencil and primary pulses with low levels of phase and intensity noise. As with conventional optics, the primary pulse must have a clean wavefront as it enters the optical assembly to ensure near diffraction-limited focusing to the far-field. While XPM is insensitive to phase noise, the induced focusing depends sensitively on the transverse intensity profile of the stencil. A noisy stencil intensity can imprint phase aberrations onto the primary pulse that inhibit its focusing and modify the far-field properties. Further, for a large stencil intensity or in a thick Kerr lens, the intensity noise can seed filamentation \cite{Yablonovitch}. To avoid unwanted, nonlinear propagation of noisy pulses, the phase acquired from XPM in the Kerr lens should be no more than a few $\pi$, i.e., $\phi_{X} = \alpha k_0n_2I_S\Delta z \lesssim 3\pi$, where $\Delta z $ is the maximum thickness of the lens \cite{fibich2005self,boyd2019nonlinear}. In practice, this amounts to limiting the focal range to a few Rayleigh ranges. The use of low-noise pulses significantly relaxes this condition: $w_0^{-2}n_2I_S(\Delta z)^2 \ll 1$, i.e., the self-focusing collapse distance must be much larger than the lens thickness \cite{marburger1975self}. Together, the conditions to avoid self-phase modulation and filamentation place an upper bound on far-field intensity of the flying focus X ($I_{f,\text{max}}$): for clean beams, $I_{f,\text{max}} \ll R^2/8n_2w_f^2$ or $I_{f,\text{max}} \ll 1\times10^{19} \,\mathrm{W/cm^2}$ for the parameters in Table I. Thus, the requirements of a low stencil pulse intensity in the near-field do not preclude a high intensity in the far-field.

\section{Summary and Conclusions}

A new technique for spatiotemporal control, the flying focus X, can create an ultrashort, arbitrary trajectory intensity peak, with any transverse profile or OAM, over distances much greater than a Rayleigh range. The technique combines cross-phase modulation in a Kerr lens with temporal pulse shaping. In a Kerr lens, a shaped auxiliary, or ``stencil," pulse imprints a second, primary pulse with a time-dependent focusing phase. In the far-field, the intensity peak formed by the time-dependent focusing of the primary pulse follows a trajectory dictated by the shape of the stencil. Depending on the shape, the peak can move forward or backwards and at any velocity. The use of a stencil pulse and XPM overcomes constraints inherent to the optical configurations employed in previous techniques for spatiotemporal control. Specifically, the flying focus X provides independent control over the shape of the transverse profile, OAM, and duration of the intensity peak. Ultimately, this additional flexibility could further enhance the utility of spatiotemporally shaped pulses for laser-based applications. 

\section*{Acknowledgments}
The authors would like to thank J.L. Shaw and K.L. Nguyen for fruitful discussions. The work published here was supported by the U.S. Department of Energy Office of Fusion Energy Sciences under contract no. DE-SC0016253, the Department of Energy under cooperative agreement no. DE-NA0003856,  the University of Rochester, and the New York State Energy Research and Development Authority. 

This report was prepared as an account of work sponsored by an agency of the U.S. Government. Neither the U.S. Government nor any agency thereof, nor any of their employees, makes any warranty, express or implied, or assumes any legal liability or responsibility for the accuracy, completeness, or usefulness of any information, apparatus, product, or process disclosed, or represents that its use would not infringe privately owned rights. Reference herein to any specific commercial product, process, or service by trade name, trademark, manufacturer, or otherwise does not necessarily constitute or imply its endorsement, recommendation, or favoring by the U.S. Government or any agency thereof.

\newpage
\section*{Disclosures}
The authors declare no conflicts of interest.

\section*{Appendix I: Kerr Lens Model}

To determine the profile of the primary pulse after the Kerr lens, consider the wave equation for the transverse electric field ($\text{E}_P$) in a time-dependent, Kerr-nonlinear medium:
\begin{equation}\tag{A1}
[\partial_t^2n(\textbf{x},t)^2-c^2\nabla^2]\text{E}_P = 0,
\end{equation}
where $n(\textbf{x},t) = n_0(\textbf{x})+\alpha n_2(\textbf{x})I_S(\tau)$ is the total refractive index, $n_0$ is the linear refractive index, $\tau = t - \int{n_0dz}/c$, and $I_S \gg I_P$ has been assumed. The spatial dependence of $n_0(\textbf{x})$ and $n_2(\textbf{x})$ indicates that the refractive index changes across the Kerr lens-vacuum interface: outside the lens $n_0=1$ and $n_2=0$. Upon switching to the coordinates $z$ and $\tau$, using $n_0\gg \alpha n_2I_S$, and noting that the lens is thin enough to neglect diffraction, Eq. (A5) becomes
\begin{equation}\tag{A2}
[c\partial_z + \alpha\partial_{\tau}n_2I_S(\tau)]\text{E}_P = 0.
\end{equation}
This equation admits analytic solutions. For the special case of $I_S(\tau) = (R/\alpha f_0n_2)(1 - f_0/f_l - u\tau/f_0)$, valid when $L_f \ll f_l$, the solution is $\text{E}_P = e^{\int \hspace{-.2em}\kappa dz}\text{E}_0[\textbf{r},\psi(\textbf{x},\tau)]$, where $\text{E}_0[\textbf{r},\tau]$ is the electric field of the primary pulse at the entrance to the Kerr lens, $\kappa = Ru/cf_0^2$, $\psi(\textbf{x},\tau) = e^{\int \hspace{-.2em}\kappa dz}(\tau-\gamma) + \gamma$, and $\gamma = (1-f_0/f_l)(f_0/u)$. The exponential factor captures the temporal broadening (compression) and the corresponding decrease (increase) in amplitude resulting from the time-dependent refractive index induced by the ramp-up (down) stencil.

The boundaries of the Kerr lens are defined by its entrance plane at $z=z_0<0$ and its exit surface: $s(r)=z_1-r^2/2R$, where $z_0<z_1<0$. The thickness and maximum radius are then $d = z_1-z_0$ and $r_{\text{max}} = (2Rd)^{1/2}$, respectively. For the Kerr lenses of interest, $\kappa d \ll 1$. As a result, the exponential factor in $\psi$ can be Taylor expanded. At the exit of the Kerr lens, i.e., $z=z_1$, $\int \hspace{-.2em}\kappa dz = \kappa(d-r^2/2R)$ and 
\begin{equation}\tag{A3}
\psi(r,\tau) = \tau + \frac{r^2}{2cf_{lK}} + \frac{\alpha n_2I_S(\tau)r^2}{2cR} - [n_{lK} + \alpha n_2I_S(\tau)]\frac{d}{c},
\end{equation}
where $f_{lK}=R/(n_{lK}-1)$ and $n_{lK}$ are the linear focal length and refractive index of the Kerr lens, respectively. A small radius of curvature enhances the nonlinear focusing, but also results in strong linear focusing that can limit the focal range ($L_f$). To overcome this, a preliminary linear lens is used to compensate the linear focal length of the Kerr lens ($f_{lK})$, while a subsequent, weaker linear lens provides additional focusing. After the optical assembly, i.e., at $z=0$,
\begin{equation}\tag{A4}
\psi(r,\tau) = \tau + \frac{r^2}{2cf_{l}} + \frac{\alpha n_2I_S(\tau)r^2}{2cR} -  \alpha n_2I_S(\tau)\frac{d}{c},
\end{equation}
where constant phase terms have been dropped. Equation (A8) along with $\text{E}_0[\textbf{r},\tau]$ define the initial condition: $\text{E}_P = e^{\kappa(d-r^2/2R) } \text{E}_0[\textbf{r},\psi(r,\tau)]$. 

The electric field of the primary pulse at the entrance to the Kerr lens can be expressed as a rapidly oscillating carrier modulating a slowly varying envelope: $\text{E}_0 = \frac{1}{2}e^{-i\omega_0\tau}E_0(\textbf{r},\tau) + \text{c.c.}$, where $\omega_0=ck_0$ is the central frequency of the pulse. Performing the substitution $\tau \rightarrow \psi(r,\tau)$, in accordance with the initial condition, illustrates that the lens phases acquired in the optical assembly appear in the exponential prefactor, while the radial group delay and dispersion appear in $E_0$. Inclusion of the radial group delay and dispersion are critical for correctly modeling the duration of the intensity peak.

\section*{Appendix II: Defocusing Kerr Lens}

A Kerr lens with a concave, quadratic shape acts as a defocusing lens. In this case, the primary pulse exits the Kerr lens with the phase
\begin{equation}\tag{A5}
    \phi_K(r,\tau)=\frac{\alpha k_0 r^2 n_2 I_S(\tau)}{2R},
\end{equation}
such that each time slice has a focal length
\begin{equation}\tag{A6}
    f(\tau)=f_l\Bigl[1-\frac{\alpha f_ln_2I_S(\tau)}{R}\Bigr]^{-1}.
\end{equation}
This exhibits a somewhat counterintuitive effect: cross-phase modulation in a self-focusing medium (i.e., one with a positive $n_2$) can induce defocusing. Here, a time slice that was overlapped with a low stencil intensity in the Kerr lens will focus closer to the optical assembly and the linear lens focal point, while a time slice that was overlapped with a high stencil intensity will focus farther from the optical assembly and the linear lens focal point. 

For the same maximum stencil intensity, a defocusing Kerr lens can produce a longer focal range than a focusing Kerr lens. The focal point of a defocusing Kerr lens is bounded by $f=f_l$ and $f = \infty$, while that of a focusing Kerr lens is bounded by $f=0$ and $f = f_l$. Specifically, the focal range for a defocusing Kerr lens is given by 
\begin{equation}\tag{A7}
    \frac{L_f}{f_l}=\frac{1}{1-\frac{\alpha f_ln_2}{R}I_{S,\mathrm{max}}} - \frac{1}{1-\frac{\alpha f_ln_2}{R}I_{S,\mathrm{min}}}, 
\end{equation}
where, in order for there to be a farthest focus, $I_{S,\mathrm{max}}<R/\alpha f_ln_2$. Note, however, that the f-number, focal spot size, and on-axis intensity of the primary pulse will vary substantially over long focal ranges ($L_f\lesssim f_l$).

To create an intensity peak that travels through the focal region at a constant velocity ($v_f$), the intensity profile of the stencil pulse must have the form
\begin{equation}\tag{A8}
I_S(\tau)=\frac{R}{\alpha f_ln_2}\Bigl(1- \frac{f_l}{f_0 + u\tau}\Bigr),
\end{equation}
where $u = cv_f/(c-v_f)$. In contrast to a focusing Kerr lens, superluminal or backwards focal velocities ($u<0$) require a stencil intensity that ramps down in time, while subluminal focal velocities ($u>0$) require a stencil intensity that ramps up in time. For focal ranges much smaller than the linear lens focal length ($L_f \ll f_l$), the stencil profile can be approximated as a simple linear ramp: $I_S(\tau) \approx (R/\alpha f_0n_2)(f_0/f_l - 1 + u\tau/f_0)$.

\section*{Appendix III: Simulation Details}
The simulations solved for the evolution of the primary pulse in three steps. The first step used the analytic model presented in Appendix I to determine the profile of the primary pulse after the optical assembly. The second step used a frequency domain Fresnel integral, as described in the Appendix of Ref. \cite{palastro2018ionization}, to propagate the envelope of the primary pulse to the far-field. The final step used the modified paraxial wave equation to propagate the primary pulse through the far-field. The results shown in Figs. 3 and 4 were obtained from this last step.

To solve for the evolution of the primary pulse after the optical assembly, $\text{E}_P$ is written as a carrier modulating an envelope: $\text{E}_P = \frac{1}{2}e^{-i\omega_0\tau}E_P(\textbf{r},z,\tau) + \text{c.c.}$, with the initial condition $E_P(\textbf{r},0,\tau) = e^{\kappa(d-r^2/2R) -i\omega_0(\psi-\tau ) }E_0[\textbf{r},\psi(r,\tau)]$. Propagation from the optical assembly to the far-field is calculated using the frequency-domain Fresnel integral: 
\begin{equation}\tag{A9}
\hat{E}_P(\textbf{r},z,\omega) = \frac{k_\omega}{2\pi i z}\int \text{exp}\Bigl[i\frac{k_\omega}{2z}(\textbf{r}-\textbf{r}^\prime)^2\Bigr]\hat{E}_P(\textbf{r}^\prime,0,\omega)d\textbf{r}^\prime,
\end{equation}
where the caret represents a Fourier transform with respect to $\omega$, i.e., the conjugate variable to $
\tau$, and $k_\omega = (\omega_0 + \omega)/c$. This allows for separate spatial resolution in the near and far-field, which greatly reduces the number of grid points compared to numerically solving the wave equation over the entire propagation distance \cite{palastro2018ionization}. In the far-field, the envelope of the primary pulse is evolved using the modified paraxial wave equation \cite{zhu2012studies}: 
\begin{equation}\tag{A10}
[2(i\omega_0-\partial_\tau)\partial_z + c\nabla_\perp^2]
E_P(\textbf{r},z,\tau) = 0.
\end{equation}
The mixed space-time derivative in Eq. (A10) ensures that effects such as radial group delay and angular dispersion are modelled correctly---a requirement for resolving the effective duration of the flying focus X. Note that after the optical assembly $\tau = t - z/c$ consistent with propagation in vacuum. 

The model described above does not account for the dispersion of $n_2$. Simulations of a 20 fs pulse using the focal geometry described in Table I predict that the dispersion of $n_2$ in ZnS increases the duration by $\approx 10\,\text{fs}$. 


\bibliography{ffx.bib}

\end{document}